# Intensity-Based Feature Selection for Near Real-Time Damage Diagnosis of Building Structures

**Seyed Omid Sajedi**

Graduate Research Assistant

Department of Civil, Structural and Environmental Engineering, University at Buffalo

Buffalo, NY, United States
*ssajedi@buffalo.edu*

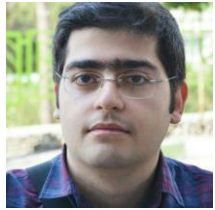

Omid is a PhD candidate interested in the autonomous structural health monitoring utilizing machine learning algorithms to reduce the recovery time after earthquakes.

**Xiao Liang**

Assistant Professor of Research

Department of Civil, Structural and Environmental Engineering, University at Buffalo

Buffalo, NY, United States
*liangx@buffalo.edu*

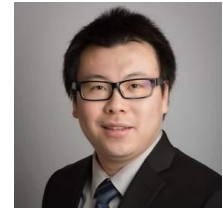

Prof. Liang specializes in performance-based methodologies for hazard resilience of buildings and infrastructure with a focus on potential applications of artificial intelligence.

**Contact:** liangx@buffalo.edu

## 1   Abstract

Near real-time damage diagnosis of building structures after extreme events (e.g., earthquakes) is of great importance in structural health monitoring. Unlike conventional methods that are usually time-consuming and require human expertise, pattern recognition algorithms have the potential to interpret sensor recordings as soon as this information is available. This paper proposes a robust framework to build a damage prediction model for building structures. Support vector machines are used to predict the existence as well as the probable location of the damage. The model is designed to consider probabilistic approaches in determining hazard intensity given the existing attenuation models in performance-based earthquake engineering. Performance of the model regarding accurate and safe predictions is enhanced using Bayesian optimization. The proposed framework is evaluated on a reinforced concrete moment frame. Targeting a selected large earthquake scenario, 6,240 nonlinear time history analyses are performed using OpenSees. Simulation results are engineered to extract low-dimensional intensity-based features that can be used as damage indicators. For the given case study, the proposed model achieves a promising accuracy of 83.1% to identify damage location, demonstrating the great potential of model capabilities.

**Keywords:** Damage Diagnosis, Hazard Resilience, Near Real-time SHM, Rapid Condition Assessment, Structural Health Monitoring

## 2 Introduction

Resilience is an important characteristic of large communities with a significant number of civil infrastructure. One of the fundamental aspects of a resilient system is the reduced time to recovery [1]. Therefore, reliable information on the condition of essential buildings such as hospitals should be available rapidly after extreme events (e.g., earthquakes). This task is difficult to achieve with manual inspections due to several limitations. For example, fast and efficient condition assessment requires monetary resources plus several teams of experts. Nonetheless, even with such resources, manual (especially visual) inspections [2-3] are mainly focused on the local damages and do not necessarily provide insight on a higher level structural performance (e.g., the story drifts). On the other hand, vibration data includes valuable information regarding the structural condition because variations in structural properties (e.g., damping and stiffness) could be indicators of damage. Moreover, installing accelerometers on buildings is an inexpensive option while the interpretation of vibration data could still be a challenging and time-consuming task which is conventionally performed utilizing system identification strategies [4].

With the development of machine learning (ML) algorithms and advances in computing technology, data-driven structural health monitoring (SHM) has shown a great potential for damage diagnosis of structures in near real-time. Given the uncertainties in extreme events and also a highly nonlinear relationship between vibration data and damage, ML can be effectively utilized to rapidly identify damage. However, reliable implementation of such frameworks depends on the proper design of the pattern recognition algorithm and tuning of hyperparameters. In this paper, a damage diagnosis framework is proposed to identify the existence and location of damage in near real-time utilizing support vector machines (SVMs).

## 3 Intensity-Based Features

Acceleration records are valuable sources of information since building response is commonly affected with the presence of damage. However, time-histories are recorded in the order of thousands of time-steps for a single sensor. This could lead to a huge computational demand due to large input sizes. Regardless of high dimensionality, the duration of earthquake excitations is not always equal as input while most data-driven models cannot adapt to variable-sized inputs. To overcome these challenges, proper selection of damage-sensitive features from vibration data is essential. Cumulative intensity measures ($I^\eta$) have been used in the earthquake engineering community for a general estimation of damage based on earthquake excitation [5-8]. $I^\eta$ is expressed as:

$$I^\eta = \int_0^{t_e} |a(t)|^\eta dt \qquad (1)$$

where $t_e$ is the duration of a ground motion (GM) record, $a$ is the acceleration in time-step $t$, and $\eta$ is a hyperparameter.

In this study, we utilize $I^\eta$ to preprocess acceleration records from the sensor placements in different degrees of freedom (DOFs) in a structure. Setting aside the sensitivity to damage, this measure provides two computational advantages. First, the size of the input is significantly reduced. Second, a unique scaler is obtained by the integration regardless of different ground motion durations ($t_e$). Beyond the computational gain, it can be shown that $I^\eta$ is correlated with damage. The ratio of $I^\eta$ values between two different sensors can be expressed as $R^\eta$. While the structure remains elastic (regardless of GM intensity), this ratio is constant. However, as damage (e.g., nonlinearity) appears in the structure, the ratio will start changing. Therefore, $R^\eta$ can be used to identify damage existence while some information regarding the intensity of GM may be lost by finding this ratio. To help avoid such loss of information, the cumulative intensity of ground motion ($I_g^\eta$) is also considered in the input. Assuming a single $\eta_i$ value, the preprocessed information from several sensors placements can be summarized as:

$$x^{\eta_i} = [I_g^{\eta_i}, R_1^{\eta_i}, R_2^{\eta_i}, \ldots, R_j^{\eta_i}] \qquad (2)$$

where index $j$ denotes the number of different combinations of two sensor placements where $R^\eta$ is calculated.

Considering the nonlinearity of mapping between input features and damage patterns, it is appropriate to augment the feature space to deal with nonlinearities. Hence, multiple $\eta$ values ($k$) are considered for a single earthquake event which yields to a stacked input vector **X** for the structure:

$$X = [x^{\eta_1}, x^{\eta_2}, \ldots, x^{\eta_k}] \quad (3)$$

## 4 Damage Diagnosis with SVMs

SVMs are among some of the advanced pattern recognition algorithms that can perform computationally efficient binary classification while showing robustness against outliers. These classifiers can build sophisticated decision boundaries to identify the existence of damage. Assuming $n_{obs}$ pairs of training observations, the constant weights $(\boldsymbol{\beta}, \beta_0)$ in an SVM classifier are calibrated to perform damage prediction in future events. Possible outcomes $(y_r)$ of observation $r$ can be defined as damage $(D)$ or no damage $(N)$. As such, these weights are determined by the following constrained optimization problem:

$$\min_{\boldsymbol{\beta}, \beta_0} \frac{1}{2} \|\boldsymbol{\beta}\|^2 + \theta_1 \theta_2 \sum_{y_r \in D} B_r \xi_r + \theta_2 \sum_{y_r \in N} B_r \xi_r$$

subject to:

$$\xi_r \geq 0, \quad y_r[h(\mathbf{X}_r)^T \boldsymbol{\beta} + \beta_0] \geq 1 - \xi_r \quad \forall r \quad (4)$$

where $\xi_r$ is the slack variable allowing a certain level of misclassifications, and $B_r$ is the normalized weight of observation $r$ with respect to its probability of occurrence compared with the other training observations. The term $h(\mathbf{X})^T \boldsymbol{\beta}$ can be further simplified utilizing the radial basis function:

$$h(\boldsymbol{X})^T \boldsymbol{\beta} = \sum_{r=1}^{n_{obs}} \alpha_r \, y_r \, exp(-\theta_3 \|\boldsymbol{X} - \boldsymbol{X}_r\|^2) \quad (5)$$

The optimization problem in (3) is solved using the dual form of Lagrange multipliers $(\alpha_r)$ [8]. Finally, a prediction function $(PRD(\mathbf{X}))$ can be developed to label future observations:

$$PRD(\boldsymbol{X}) = \begin{cases} D & h(\boldsymbol{X})^T \boldsymbol{\beta} + \beta_0 > 0 \\ N & else \end{cases} \quad (6)$$

In the SVM formulation, the three hyperparameters $\theta_1, \theta_2, \theta_3$ are used to adjust the performance of the classifier. For the case of damage existence, a single SVM classifier is sufficient to predict the existence of damage in a building. However, information about the location of damage can also be critical in SHM. A system of SVM classifiers can be developed to locate defects in structures such that multiple damaged components can be detected simultaneously. In this case, one SVM classifier is considered for each potential damage location. The binary output of the SVM at each story will correspond to the existence of damage in that location. This combination of binary outputs can be utilized as an extension of such classifiers for cases where several potential damage locations exist.

The described process can be used to build a framework to identify the existence and also the location of damage in building structures. At first glance, this may seem a straightforward task; however, the machine learning algorithms have certain hyperparameters that affect the performance. For example, increasing $\theta_1$ will put more emphasis on the correct prediction of observations associated with the damage. Moreover, SVMs are capable of constructing highly nonlinear decision boundaries which may result in overfitting. The classifier could output nearly perfect predictions for the training data but lose generalization for future events. To avoid this problem, $\theta_1, \theta_2, \theta_3$ and $\eta$ values are selected by Bayesian optimization [9] considering 10-fold cross-validation [10]. The simplest predefined objective function can be the global accuracy ($GA$). However, there is commonly a significant imbalance between damaged and undamaged components in building structures. For example, instability in moment frames could be dominated by a soft-story mechanism where the majority of structural members are not locally damaged. In this case, maximizing the global accuracy could possibly lead to a classifier that labels all locations as $N$ even if there is damage. This is obviously not acceptable in SHM, the correct prediction of $D$ class is of greater importance compared with $N$. Moreover, it is of interest to model the uncertainties from observations in the training process which requires a score quantity:

$$s_{ij} = \sum_{r=1}^{n_{obs}} P_r I_{ij}^r \quad (7)$$

where $P_r$ is the probability of occurrence for observation $r$ and $I_{ij}^r$ is an identity function which is equal to one if that observation is predicted as class $j$ but has the ground truth label $i$, and zero otherwise. Assuming that there could be $p$ different possible classes sorted based on severity, a cost objective function is expressed as:

$$C = -w_1 \sum_{m=1}^{p} s_{mm}$$
$$+ w_2 \sum_{m=2}^{p} \sum_{n=1}^{m-1} \Omega_n \lambda_{mn} s_{mn}$$
$$+ w_3 \sum_{n=2}^{p} \sum_{m=1}^{n-1} \Omega_n \lambda_{mn} s_{mn} \quad (8)$$

where $\lambda_{mn}$ is a misclassification penalty factor for the corresponding score value. The first term in (8) indicates the accurate predictions while the second and third one, respectively, correspond to the underestimated and conservative estimations. $\Omega_n$ is a correction factor that can be set to a large number for nonexistent classes in the training data. The purpose of this correction factor is to exclude predictions that do not exist in the training set. $w_1, w_2,$ and $w_3$ are predefined weights that are selected to balance the trade-off between the three constituting terms in (8). It is more likely for a decision maker to favor conservative predictions over underestimation of damage ($w_2 > w_3$). By minimizing $C$, SVM hyperparameters are optimized to form a decision boundary that deals with uncertainties from observations and also consequences of misclassification.

## 5 Damage Diagnosis of a 3D RC frame

The proposed damage identification framework mentioned earlier is investigated on a 3D reinforced concrete moment frame. The finite element model is created in OpenSees [11], assuming that each floor acts as a rigid diaphragm. Acceleration is recorded for the three floors and the ground level in the two principle directions as shown in Figure 1. To calculate $R^\eta$, the ratio between top and bottom sensor recordings for the three stories are separately considered in the two directions.

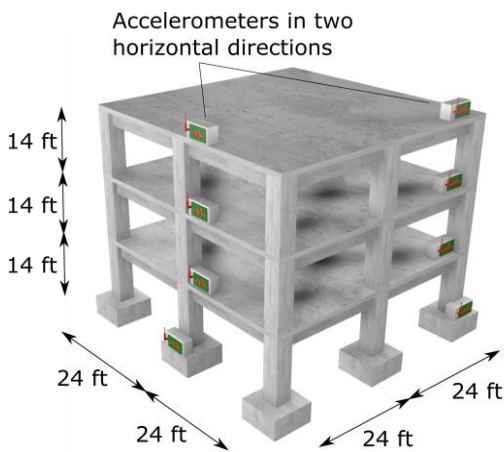

*Figure 1. 3D RC frame geometry and sensor placements*

To model damage, an M7 earthquake scenario is selected which yields 208 GMs [12]. The uncertainty of seismic events for 30 different scale factors is obtained from the CB-14 attenuation model [13]. 6,240 nonlinear time-history analyses [14] are performed. The immediate occupancy criteria by FEMA 356 (exceeding 0.5% peak inter-story drift) is considered as damage [15]. Two different models are considered: the existence and the location of damage. For the first one, the most critical peak story drift ratio is considered to label the whole building as damaged or not. For the damage location model, class IDs are identified with three letters. For example, DDN indicates the number of observations that there exist damage in the first and second stories but no damage in the third one.

As mentioned earlier, a system of multiple SVM classifiers is required to identify the location of the damage. Mathematically speaking, $2^3$ different patterns are possible. For the set of selected earthquakes and taken into account the specific physical and dynamic properties of the building, certain damage mechanisms do not exist in the datasets (e.g. NND). To avoid such predictions by SVMs, a large penalty factor $\Omega = 100$ is set for such predictions in evaluating the cost function in (8). Consequences of misclassification are included in the cost function by assuming $\lambda_{mn} = |m - n|^2$. It should be noted that for each model, different number of $\eta$ features ($k$) are investigated.

### 5.1 Damage Existence Model

In this model, four different combinations of $w_1, w_2,$ and $w_3$ are studied. The hyperparameters are individually optimized for each one as shown in Figure 2. The prior assumption of balancing weights ($w$) in the cost function can affect the performance indicating that there is a trade-off between accuracy and how conservative the model's predictions are. To better investigate the performance with respect to different classes, confusion score matrices (Figure 2) are utilized where rows and columns, respectively, correspond to the ground truth and predicted labels. Furthermore, numerical score values ($s_{ij}$) as in (7) are calculated and presented in a normalized manner (for each row) to observe the class accuracies as in Figure 2. It can be observed that a decision-maker can change the balance among the terms in the cost function to obtain a classifier which is more accurate to predict damage but also more conservative as in Figure 2(d). Score values are also indicators that the structure remained undamaged for a larger portion of dataset as the sum of score values for the first row (N) is relatively larger.

## 5.2 Damage Location Model

A similar procedure is repeated while the number of hyperparameters for the SVMs is tripled due to the fact that each story has one classier. The prediction outcome is used to minimize a single cost objective function as in (8). The confusion matrix for the building is expressed in Figure 3. It can be seen that compared with the first model, although the total number of events is constant, there are less training observations for each damage pattern. Moreover, the imbalance between classes in more significant. Overall, the model shows promising robustness with 83.1% global accuracy. One may also note that the sum of lower diagonal score values is significantly smaller than the upper diagonal (conservative predictions), showing that the model is optimized to minimize underestimation of damage.

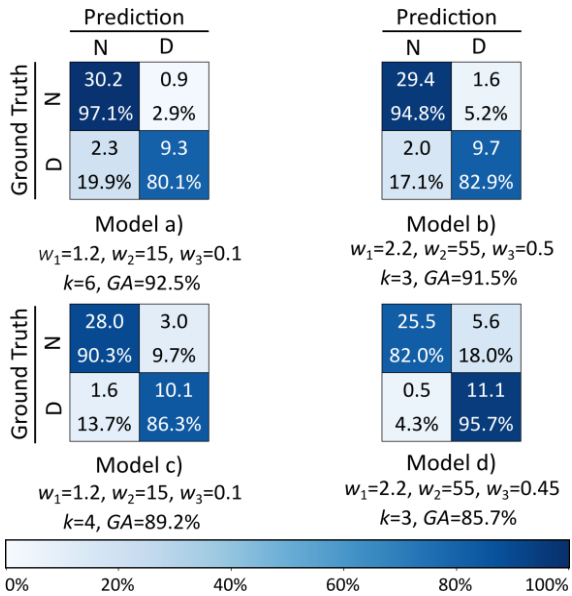

Figure 2. Test results for the damage existence model.

|  |  | Prediction | | | |
|---|---|---|---|---|---|
|  |  | NNN | DNN | DDN | DDD |
| Ground Truth | NNN | 26.51 / 85.5% | 4.49 / 14.4% | 0.02 / 0.1% | 0 / 0% |
| | DNN | 1.38 / 14.6% | 7.64 / 81.1% | 0.38 / 4.0% | 0.02 / 0.3% |
| | DDN | 0 / 0.0% | 0.66 / 31.7% | 1.17 / 56.3% | 0.25 / 12.0% |
| | DDD | 0 / 0.0% | 0 / 0.0% | 0 / 0.0% | 0.16 / 100% |

Figure 3. Test results for the damage location model ($w_1$=12, $w_2$=5, $w_3$=0.05, k=6, GA= 83.1%).

## 6 Conclusions

This paper focuses on a data-driven SHM technique utilizing SVMs. Considering the capability of these classifiers in constructing complex decision boundaries, a cost-sensitive objective function was designed to model uncertainties in observations and also the consequences of misclassifications. Sensitive hyperparameters in SVM and input data were optimized to obtain enhanced performance. The proposed framework shows a promising potential to perform near real-time damage diagnosis given its robustness and relatively inexpensive cost of computation.